\documentstyle[leqno,epsfig]{article}
\begin{document}
\renewcommand{\baselinestretch}{1.15}
\large

\centerline{\bf Modeling the effects of tetanus vaccination}
\centerline{\bf on chronically infected HIV patients} 
\bigskip
\centerline{L. E. Jones and A. S. Perelson}
\centerline{Theoretical Biology and Biophysics Group}
\centerline{Los Alamos National Laboratory}
\centerline{Los Alamos, NM 87545}
\medskip
\bigskip
\smallskip
\begin{flushleft}
Running Head: Modeling effects of vaccination\\
\bigskip
Corresponding author: Alan S. Perelson, Theoretical Biology and Biophysics, MS-K710,
Los Alamos National Laboratory, Los Alamos, NM 87545, asp@lanl.gov\\
\bigskip

Requests for reprints: Alan S. Perelson, Theoretical Biology and Biophysics,
MS-K710, Los Alamos National Laboratory, Los Alamos NM 87545, (505)667-6829,
asp@lanl.gov\\

\bigskip
\bigskip
This work was performed under the auspices of the U.S.
Department of Energy and was supported by NIH grants
AI28433 and  RR06555.
\end{flushleft}

\pagebreak
\noindent
{\bf Objective:}
To model the effects of vaccination with a common recall
antigen on chronically infected HIV-1 patients.

\noindent
{\bf Background:}
T-cell activation plays a critical role in the initiation and
propagation of HIV-1 infection and yet
transient activation of the immune system is a normal response to
immunization. While it is now considered wise to vaccinate
HIV-1 positive patients, it is crucial to anticipate any lasting
effects of vaccination on plasma HIV-1 RNA levels and on infected T
cell populations.

\noindent
{\bf Method:}
We extend a simple dynamical model of HIV infection
to include T cell activation by vaccination. We show that
the model can reproduce many but not all of the
features of the post-tetanus immunization rise in viral load
observed and reported on by Stanley et al. [{\em N. Engl.
J. Med.}, 334:1222--1230 (1996)].

\noindent
{\bf Results and conclusions:}
Amplitudes and approximate timing of
post immunization peak viral loads 
were matched in nine of eleven cases; in patients with double
post immunization peaks of nearly equal amplitude the later peaks were matched.
Furthermore, our simulations suggest that productively infected cell
populations track post-vaccination increases in plasma viral load, rising and falling
in concert on a period of about 4 weeks, while chronically infected cells peak later and
remain elevated over baseline levels for up to six weeks post-vaccination.

\noindent
{\bf Keywords: chronic infection, mathematical models, plasma viral load} 

\pagebreak
\pagestyle{plain}
\pagenumbering{arabic}

\section*{Introduction}
T-cell activation plays a critical role
in HIV infection and progression to AIDS [1--4].
In vitro studies have demonstrated the
importance of cell activation in establishing productive
HIV infection.  For example,
reverse transcription within resting cells
may be incomplete and integration of proviral DNA may not
occur, resulting in abortive infection [5--6].
In contrast, activated
peripheral blood mononuclear cells are readily infected in
culture [5--7].
The course of HIV infection in vivo appears to be influenced by
cell activation. High levels of activated peripheral T cells
are a predictor of early progression to AIDS [8--9],
and chronic immune activation due
to frequent and repeated parasitic infection has been suggested
as the probable cause of the increased rate of HIV progression to
AIDS and the
greater susceptibility to infection observed in sub-Saharan Africa
as compared to the developed countries [10--12].

Transient activation of the immune system occurs during
infections and is a normal response to
immunization.  Such activation can affect the course of HIV infection.
Numerous studies have examined the consequences of vaccinating
untreated HIV-infected individuals [2, 13--17].
Stanley et al. [2] found that giving a booster dose of tetanus toxoid resulted in
transient increases in plasma viremia in all 13 HIV-infected subjects
studied. Staprans et al. [13] found that influenza vaccination 
led to transient increases of plasma HIV-1 RNA, and that
patients with higher CD4 T-cell counts had larger and more rapid
increases in viral load. 
The large immunization-related increases in virus reported by 
Staprans et al. [13]
suggest that the observed viral replication may be correlated  with activation of
both antigen-specific T-cells and other non-specific T-cell populations, which
once activated, then become targets for infection [3].

Other studies using influenza vaccination and pneumococcal vaccine
have confirmed that vaccination can trigger increases in plasma
HIV RNA [14--17].
Thus a number of data sets are available that
illustrate a relationship between immune system activation and changes in 
HIV viral load. This data presents an interesting challenge to current
mathematical models of HIV infection [18--20].
In an early theoretical paper, McLean and Nowak
[21] examined the  effects of infection with pathogens
other than HIV as a cofactor in HIV progression. However, no quantitative
data was available at the time for direct comparison of model and
theory.  A more recent study [22]
presented a model and some comparisons with
averaged data for untreated patients, but concentrated primarily on 
antigen-driven T-cell proliferation and residual viral replication
for patients on anti-retroviral therapy. Here we focus on the effects of
vaccination on untreated HIV-infected patients.  We
construct a simple  model of HIV-infection and immune system activation
via vaccination, and then compare the model with data from the study
of Stanley et al. [2] in which 13 HIV seropositive, asymptomatic
individuals were vaccinated with tetanus toxoid.

\section*{A simple vaccination model}

To  understand the effects of vaccination
with a common recall antigen (tetanus toxoid)
on chronically infected, untreated HIV patients whose viral
loads have reached steady state, we modify what has become
a  standard HIV infection model [18-19].
The model includes uninfected, infected and chronically 
infected T-cell populations, HIV and other antigens $A$.
We assume that the vaccine--introduced antigens, $A$,
activate CD4+ T-cells. For simplicity, the model ignores latently
infected cells, which in untreated individuals are an extremely
minor source of virus;
Chun and Siliciano [23] estimate that only 1 in
$10^5$ T cells are latently infected. We also
do not separately consider vaccine-specific or  HIV specific T-cell
populations, since this leads to a model with more parameters
than the data reported by Stanley et al. [2] will support.

The model, like that of McLean and Nowak [21],  includes
vaccine antigen $A$, which we assume is eliminated from the
body at a rate proportional to both the antigen concentration
and the CD4+ T cell density.  The rate of proportionality
$\gamma$ implicitly includes a factor that accounts for the fact
that only a fraction of T cells are antigen specific and accounts for
complexities of the antigen specific, helper cell dependent
response that leads to antigen elimination. 

Since individuals would be
expected to vary with regard to the time of their last tetanus
vaccination, one might assume that levels of antibody and T memory cells
 specific for tetanus toxoid would also vary. In our model
we avoid this level of detail, but simply assume that
the antigen clearance rate constant, $\gamma$, and parameters
that determine the rate of T cell activation vary for each patient. 

The model we use is given by the following system of differential
equations:

\begin{eqnarray}
\frac{dA}{dt} &  = & - \gamma AT \\ 
\frac{dT}{dt} &  = & \lambda + a(\frac{A}{A + K})T - dT - kVT  \\
\frac{dT^*}{dt} & =  & (1 - \alpha)kVT - \delta T^*  \\
\frac{dC}{dt} & = & \alpha kVT - \mu C \\
\frac{dV}{dt} & = &  N\delta T^* + N_c\mu C - cV  
\end{eqnarray}

\noindent where $A$ is the vaccine antigen, $T$ are uninfected CD4+ T 
cells, $T^*$ are productively infected cells, $C$ are
chronically infected cells, and $V$ represents HIV.  Vaccine
antigen is cleared in a T cell-dependent manner with rate
constant $\gamma$.
Uninfected T cells, $T$, are produced at a rate $\lambda$,
die at a rate $d$, and are infected by virus with rate constant 
$k$.
In the presence of antigen, we assume T cells are activated into
proliferation at a maximum rate $a$, and that the proliferation
rate depends on the antigen concentration with a half-saturation
constant $K$.  $K$ is thus the antigen concentration that
drives T cell proliferation to half its maximal value. 
Productively infected cells, $T^*$, are generated by infection of
target T-cells, $T$, at a
rate $kVT$, and die at rate $\delta$, as in the standard model
[18].
Chronically infected T-cells, $C$, are produced from 
susceptible T-cells
at a rate $\alpha kVT$, where $\alpha<1$. Thus, the production of
chronically infected cells occurs at a fraction of the
rate productively infected cells are generated. Chronically infected cells die at a rate
$\mu$, which we assume is less than $\delta$, so that chronically
infected cells are longer-lived than productively infected cells. 
Free virus, $V$, is produced by productively infected
cells at average rate $N\delta$,  by chronically infected
cells at rate $N_c \mu$, and is cleared at rate $c$ per virion.
Based on previous work we assumed $\lambda = 1\times 10^4 ml^{-1}$ [24];
$d = 0.01 d^{-1}$ [25];
$\delta =  0.7 d^{-1}$ [18];
$\alpha = 0.195$ [24];
$\mu =  0.07 d^{-1}$ [19];
$N_c =  4.11$  [24], 
and $c =  13  d^{-1}$ [22, 26].
Prior to vaccination, each patient was
assumed to be in steady state with a known total baseline T
cell count, $\bar{T_T}$, and viral load, $\bar{V}$. 

Given these measurements and the steady-state conditions
derived from equations (1) -- (5) (see Appendix), 
initial numbers of infected cells can be derived as
well as a set of parameters describing chronic HIV infection
for each patient.

At time $t=0$, we assume a dose of antigen, $A_0$, was given in a
vaccine, which perturbed the steady state.   We assume the
same dose of vaccine was given to all individuals, and then
model the antigen remaining in each individual as a
fraction of the immunizing dose.  This is equivalent to setting
$A_0=1$. We numerically solve the system of differential
equations given by (1) - (5), with the initial conditions
$A(0)=1$, and $T$, $T^*$, $C$, and $V$ set to their steady state
values. For each individual studied, 
the parameters $K$, $a$ and $\gamma$ were allowed
to vary  so that  the amplitude and timing  of the response
could be best matched to the observed post-vaccination viremia.

\section*{Data and sampling}
Stanley {\em et al.} [2] studied 16 asymptomatic homosexual men
seropositive for HIV-1.
Thirteen subjects were given an 0.5 ml tetanus booster intramuscularly
and three were mock immunized. After vaccination, viral load
measurements were taken 
on days 0 (baseline), 4, 7, 14, 21, 28 and 42.
 No information on prior vaccination history, i.e.,
date of last tetanus booster, or general quality of health was given.
It was noted that one patient developed cavitary pneumonia  about a
month following vaccination.

Of the thirteen patients reported on by Stanley {\em et al.} [2], 
we did not include patients 12 and 13 in our study. Published
data for patient 12 had inconsistencies, and patient 13
had a T-cell count, $T_T$, of 8 cells per $mm^3$, which is too low
to be consistent with our parameter assumptions and
the pre-vaccination steady state assumption of our
model (see Appendix). 
Thus either the patient was not at true steady state, or the parameter values
chosen from the literature are not applicable to this patient.
The baseline characteristics for the remaining eleven patients are given
in Table 1. 

Data obtained for a subset of the patients
are shown in Figures 1a and 1b.
The data sampling was too sparse to determine the exact
peak of viremia or the time the peak was attained. However,
Stanley {\em et al.}[2] report the apparent peak and the
time it was measured.
Many patients had {\em two} post-immunization  peaks
in viral load. The larger peak was 
the earlier of the two observed peaks in patients 3, 5, 10, and
the second  of two peaks in patients 1, 4, 7, and 11. 
Patients 2, 8, and 9 had two peaks of nearly equal amplitude (Figure 1a). 

Of the patients with double post-immunization peaks in viremia,
patients 1, 7, 8, 9 and 11 had peaks at days 7 and 21 (Figure 1a),
though in some of these cases the earlier peak can be classified
as a minor peak or an inflection point. Of those with single
observed post-immunization peaks,  patients 3, 5, and 10 
have primary peaks at day 7, and inflection points, shoulders, or
minor peaks at day 28 (Figure 1b).

\begin{table}
\begin{center}
\small
\centerline{Table 1. Baseline characteristics and post-immunization
plasma viremia}
   \begin{tabular}{cccccccc}
   \hline
Patient & T cells, $T_T$ & $\bar{T}$ & $\bar{T^*}$ & $\bar{C}$ & $V_0$ $\dagger$ & $V_p$ & $V_p/V_0$\\
    & cells/$mm^3$ & cells/$mm^3$  & cells/$mm^3$ & cells/$mm^3$ & RNA/ml & RNA/ml & \\
   \hline
1. & 362 & 336 & 7.6 & 18.5 & 147,000 & 437,000 (21/20) $\ddagger$ & 2.97 \\
2. & 271 & 241 & 8.7 & 21.1 & 3,850 & 10,500 (14,28/14) & 2.73 \\
3. & 350 & 323 & 7.8 & 18.8 & 120,000 & 900,000 (7/8)  & 7.50 \\
4. & 389 & 364 & 7.3 & 17.7 & 100,000 & 700,000 (3,21/18) & 7.00 \\ 
5. & 586 & 569 & 4.9 & 12.0 &  21,000 &  45,000 (3,21/19) & 2.14  \\
6. & 336 & 309 & 7.9 & 19.3 & 215,000 & 725,000 (7/9) & 3.37 \\
7. & 336 & 309 & 7.9 & 19.3 & 75,000 & 315,000 (7,21/18) & 4.20 \\
8. & 361 & 335 & 7.6 & 18.5 & 87,500 & 241,000 (7,21/19) & 2.75 \\
9. & 497 & 476 & 6.0 & 14.6 & 220,000 & 745,000 (7,21/23) & 3.38  \\
10. & 615 & 599 & 4.6 & 11.1 & 10,500 & 375,000 (7/7) & 35.7  \\
11. & 363 & 337 & 7.6 & 18.5 & 80,000 & 225,000 (7,21/18) & 2.81 \\ 
\hline
\end{tabular}
\end{center}
\centerline{\small $\dagger$ $V_0 =$ baseline viral load, $V_p =$ peak viremia }
\centerline{\small $\ddagger$ First number(s) are observed time(s); second number, modeled time to peak viremia (d).}
\end{table}

\section*{Results}
For each of the  eleven patients studied, the model was fit to the data
on viral load changes after vaccination. Using nonlinear regression
techniques, the best-fit set of parameters were then determined
for each patient (Table 2).

The amplitude and the approximate timing of the
post-immunization maximum viremia
were matched in nine of eleven cases; in patients with double
peaks of nearly equal amplitude (i.e., patients 8, 9),
the later peaks were matched.  
The results  are 
summarized in Tables 1 and 2. 

\subsection*{Early peaks in post-vaccination plasma viral load}
Patients with relatively high baseline CD4 T-cell counts
(patients 5, 9, 10), 
as well as those with high baseline viremia (patients 6, 9) 
all had early peak viremia, at either 3 days (patient 5) or 7 days
(patients 6, 9, 10) post-immunization. This is reasonable since
patients with a high baseline T-cell count or high viral load 
 would be expected to have a larger population of
target cells that could respond 
post-immunization. Thus, a vigorous early response by the immune system
ensures an early peak in viremia.

The  ``viremia factor",  or 
ratio of peak viremia to baseline viremia, given as $V_p/V_0$ in
Table 1, 
was generally rather low for most patients, ranging between 2 and
4 (mean = 2.92) 
for nine of twelve patients. Three remaining patients with
unusually high ``viremia factor" ratios, 
patients 3 ($V_p/V_0 = 7.5$), 4 ($V_p/V_0 = 7.0$) and 10 ($V_p/V_0 = 35.7$)
also had early peaks, though in patient 4 this was a minor 
inflection followed by peak viremia at 21 days, 
whereas both patients 3 and 10 had single early peaks.  

\begin{table}
\begin{center}
\small
\centerline{Table 2. Best-fit parameter values for modeling
post-immunization plasma viremia}
   \begin{tabular}{cccccc}
   \hline
Patient No. & $k$ & $N$ & $a$ & $K$ & $\gamma$ \\
            & $ml/d$   &  & $d^{-1}$ &  & $ml/d$ \\
   \hline
1. & $1.3452 \times 10^{-7}$ & 356 &  3.66 & 187.1  & $3.8199 \times 10^{-8}$ \\
2. & $8.1743 \times 10^{-6}$ & 7  & 1.8 & 41.1 & $4.2732 \times 10^{-7}$ \\
3. & $1.7437 \times 10^{-7}$ & 285 & 1.60  & {\em 3.36} & $1.7475 \times 10^{-6}$ \\
4. & $1.7475 \times 10^{-7}$ & 253 & 1.02  &  31.3  &  $1.0001 \times 10^{-8}$ \\
5. & $3.6064 \times 10^{-7}$ & 78  &  0.78 &  {\em 3.68} & $6.6248 \times 10^{-6}$  \\
6. & $1.0411 \times 10^{-7}$ & 501 &  2.49 & {\em 2.20} & $1.0258 \times 10^{-5}$ \\
7. & $2.9845 \times 10^{-7}$ & 174 &  3.63 &  169.0 & $1.5195 \times 10^{-8}$ \\
8. & $2.2705 \times 10^{-7}$ & 211 &  2.00 & 95.2 & $6.4431 \times 10^{-8}$ \\
9. & $4.9959 \times 10^{-8}$ & 678 &  1.96 & 115.2 & $9.1986 \times 10^{-8}$ \\
10. & $6.3697 \times 10^{-7}$ &  41 &  1.75 & {\em 0.39} & $3.08277 \times 10^{-6}$  \\
11. & $2.4603 \times 10^{-7}$ & 194 &  1.91 & 86.3 & $1.0660 \times 10^{-7}$ \\
   \hline
   \end{tabular}
\end{center}
\end{table}

We were able to match well the amplitude and timing of peak viremia 
for patients 3, 6, and 10 (Figure 2), all of whom had
early peaks in viremia
and were best-fit with relatively high antigen clearance rate 
constants, $\gamma$, in our simulations (Table 2).

\subsection*{Double peaks in post-vaccination plasma viral load}
Over half of the patients studied had multiple peaks in viremia,
though often the first peak might be classified as `minor'.
For patients with a minor early peak, usually at day 7, 
followed by a maximum in viral load at day 21, we were
able to fit the amplitude and timing of the maximum peak. 
Patients 1, 4, 7, and 11 exemplify this behavior; 
results for patient 1 are shown in Figure 2.
These patients were best-fit with $\gamma$ values
that were low relative to those given for
patients with true early peaks in viremia (Table 2).
Patient 5 had very early viremia [day 4] followed by
a second peak at day 21. We were able to model the second
peak, and from this obtained a relatively high antigen clearance constant.

Some patients experienced early and late
peaks in viremia of equal or nearly equal amplitude.
Of the patients with double peaks in viremia, our model
generally fit the
later peak at day 21. These patients were assigned much lower
clearance rates then were the patients who experienced single early
peaks in viremia.  However, in patients with double peaks 
(e.g., patients 8 and 9, Figure 2)
fits to the earlier peaks, unobtainable with this model, might
require higher values of $\gamma$.

\subsection*{T-cell - Antigen Interactions}

The  rate of antigen induced  T-cell activation/proliferation 
is given in the model by
$a(\frac{A}{A + K}) $ [equations 2 and 7].   Recall that the
antigen concentration has been normalized, so that
$A_0 =1$ and $A \le 1$. Thus, the initial rate of interaction,
which is the highest rate since the antigen has not yet decayed, is
given by $a(\frac{1}{1 + K}) $  
For patients 1, 2, 4, 7, 8, 9, and 11, $K >> 1$ (Table 2). 
Thus, the denominator in the T cell activation term
is approximately equal to $K$, and the initial activation/proliferation
rate is $\sim \frac{a}{K} $. 

Substituting in $K$ and $a$ values
(Table 2) for these patients yields initial $a/K$ values ranging
from 0.017 d$^{-1}$ to 0.043 d$^{-1}$ (mean value $\pm$ sample standard deviation
 $0.025 \pm 0.0095$ d$^{-1}$) with
most values clustered around 0.02 d$^{-1}$.
These patients all had low values for the T-cell
activation/proliferation
term, and late peaks (day 21) in observed and modeled viremia.

Higher initial values of the activation term would increase the
interaction between T-cells and antigen A, resulting in swifter activation
of T-cells and an earlier increase in virus. For patients 3, 5, 6,
and 10, $A_0$ and $K$
were the same order of magnitude or within one order of magnitude
in value, and the approximation for the interaction term
shown above does not hold. In this
case, the initial interaction term for these patients ranged from $0.21$
d$^{-1}$ to $4.5$ d$^{-1}$ (mean value $1.6 \pm 1.9$), 
which is between one and two orders of magnitude greater
than the interaction terms found for the prior group of `late peaking'
patients. Patients 3, 6, and 10 all had early viremia, and
relatively high values for the antigen clearance term $\gamma$.
Note that patient 5 also had an early peak in observed viremia
at day 4 that we could not fit with this model. Patient 5
had a second peak in viral load which we could fit,
obtaining both a high initial interaction rate
and a high antigen clearance rate.  In patients 5 and 10, the
large interaction term and antigen clearance rates may reflect
relatively high baseline T-cell levels.

The parameter $a$ in our model represents the maximum rate of antigen--driven
T-cell proliferation. From Table 2, the mean $\pm$ sample standard
deviation of $a = 2.1\pm 0.91$.
For a cell population dividing at rate $a$, $1/a$ (days) is the average division time. Using the
parameter estimates from Table 2, one can compute the mean of $1/a$, and find
that the mean value of average division times is $.59 \pm 0.30$ (days) or $14 \pm 7.1$ (hours).

\subsection*{Productively and Chronically infected cells}

Of concern for HIV-infected patients are any potentially long-lasting
effects, aside from protective immunity to the vaccine antigen, associated
with immunization. According to our model, the density of productively infected cells, $T^*$,
tracks changes in viral load, peaking and falling at the same time. The effects of
vaccination on longer lived chronically infected cells, $C$, however, are
more lingering. 
According to our model, the ratio of maximum chronically infected cell level over
baseline  for all but one patient ranged from 1.6 to 3.6, with a mean of
$3.4 \pm 0.22$, and the one outlier value of a 13--fold increase in chronically
infected cells [patient 10]. At 28 days post
immunization, our simulations show these long-lived cells still elevated with respect
to their initial (steady state) values, 
yielding increases of 1.4 to 3.0 over baseline, with a mean 
of $1.9 \pm 0.51$, disregarding the outlier value which is still high
at 4.3 above baseline. At 42 days post immunization,
chronically infected cells are still a factor of $1.3 \pm 0.46$ on average
above baseline, again disregarding the outlier value of 1.9
above baseline. Indeed, focusing only on late-peaking patients,
for whom infected and chronically infected cells are later to reach maximum,
yields ratios of $2.0 \pm 0.22$ at maximum, $2.0 \pm 0.28$ at 28 days,
and $1.4 \pm 0.22$ at 42 days.
By contrast, at 42 days post immunization, 
productively infected cells for both patient populations have achieved
or fallen below
their baseline values. Modeled infected cell densities from patient 10
are shown plotted against plasma viral load (Figure 3);
note that chronically infected
cells remain elevated long after infected cell populations have
fallen to baseline values even in this early peaking patient.
Thus while vaccination gives rise to short-term
changes in viral load and productively infected cells, it
may also produce longer lasting
increases in the populations of chronically infected cells.

\section*{Discussion}

With a simple modification of a standard HIV-infection model
that takes into account antigen-driven proliferation of T cells,
we are able to reproduce the 
general features of the post-vaccination rise in viral load
reported on by Stanley et al. [2]. 
This implies that
simply increasing the number of
cells susceptible to infection by antigen activation, i.e.,
vaccination, can account for the modest rises
in viral load observed by Stanley et al. [2] after tetanus
booster vaccination. 
Interestingly,
more than half of the patients in this study
had double peaks in plasma HIV-1 RNA post-vaccination. While
assay variability might
account for some of these peaks, in other cases the peaks were
sufficiently large that this seems unlikely.  For patients 
with double peaks we were only able to
fit the later of the two peaks, which was often the larger
peak. 
These patients were assigned lower
pathogen clearance rate constants, $\gamma$, than patients with early
single peaks in viremia, which we also successfully modeled.

The fact that we could not approximate the earlier peaks in
those patients with double peaks in viremia suggests that
our model may lack some features present in the actual biology.
For example, individuals vaccinated recently may have a larger
and more robust memory cell response than individuals vaccinated
a long time in the past.  The two peaks may represent an early
memory response followed by a naive cell response.  In
individuals with little remaining memory, one may speculate that
only the later naive response would be observed.
However, since the two peaks frequently occur 14 days apart, this
explanation would require a substantial difference in the time to
activate naive and memory cell responses.
Another possible explanation for double peaks is that the first
peak is due to stimulation and infection of tetanus specific T
cells, whereas the second peak is caused by the rise in
viremia stimulating HIV specific T cells, or increasing the
level of immune system activation, which in turn leads to an 
increase in target cell availability.  We do not favor this
explanation because our model, which follows the activation of T
cells by antigen, had difficulty mimicking the fast rise in
viremia needed to explain an early peak.  One can speculate then
 that the early peak is not due to the infection of
vaccine-induced target cells, but rather to an effect
such as the cytokine-induced
enhancement  of HIV transcription and release from infected cells 
as has been seen in vitro with the use of tumor necrosis 
factor-$\alpha$  [5].

Despite the fact that the model failed to reproduce double peaks
in viral load, the model could account for the timing and
amplitude of the major peak in post-vaccination plasma viremia.
Because HIV-1, as observed in plasma, is produced by infected
cells, the model suggests that levels of both productively
infected and chronically infected cells also increase due to
vaccination.  Levels of latently infected cells presumably also
increase, but in untreated individuals this population makes a
minor contribution to plasma viral load.  According to the model
the increase in productively infected cells mirrored the
observed changes in plasma viral load, and due to the short
lifespan of these cells led to little long-term perturbation of
this population.  However, chronically infected cells have longer
lifetimes and hence vaccination led to a larger period of increase
in this population.  Nevertheless, in all but one patient the increase was not large,
with the maximum increase being 3.6--fold, and in the most extreme case
studied here, the increase was 13--fold. However, by day 42 the
chronically infected cell elevations were less than 2-fold in all
patients. This,  coupled with the
fact that chronically infected cells generally produce only
1-7\% of plasma virus [19],
implies that the long-term impact of this change on
 disease progression should be small. 
\bigskip

\leftline{\bf Acknowledgements}
We thank Rob J. DeBoer for helpful discussions.

\pagebreak
\section*{Appendix}

The steady state productively infected and chronically infected cell 
concentrations, $\bar{T^*}$, and $\bar{C}$, the infection rate
constant, $k$, and burst size, $N$, may be calculated 
from the following steady state equations, derived by setting the
left-hand sides of equations (2)-(5) to zero.

\begin{eqnarray}
k    & = & \frac{\lambda - d \bar{T}}{\bar{V}\bar{T}} \\
\bar{T^*} & = & \frac{1}{\delta}(1 - \alpha)(\lambda - d \bar{T}) \\
\bar{C} & = & \frac{\alpha}{\mu}(\lambda - d \bar{T}) \\
N & = & \frac{c \bar{V} - N_c \mu \bar{C}}{ \delta \bar{T^*}} 
\end{eqnarray}

\noindent In addition, we impose the following conservation rule:

\begin{equation}
\bar{T_T}   =  \bar{T} + \bar{T^*} + \bar{C} 
\end{equation}

\noindent where $\bar{T_T}$ is the observed (total) T-cell count, comprising
uninfected cells $\bar{T}$, productively infected cells $\bar{T^*}$,
and chronically infected cells $\bar{C}$. From the above condition
and equations (7) and (8), we derive the following relation for
$\bar{T}$:

\begin{equation}
\bar{T}  =  \frac{\delta\mu\bar{T_T} - \lambda [(1 - \alpha)\mu + \alpha\delta]}{\delta\mu - d[(1 - \alpha)\mu + \alpha\delta]}
\end{equation}

\noindent Parameters other than $k$ and $N$ were set as indicated
in the text. Note that with these parameter choices, and the measured baseline T cell
and viral load levels in the patients studied by Stanley et 
al. [2] this model generally yields larger
steady state levels of chronically infected cells than
of productively infected cells in these untreated patients (Table 2).

Note that according to equation (11), the T cell count at steady state
has to be larger than a minimal value for the numerator   
of equation (11) to be positive. For patient 13, with a T cell count of
8 cells per $\mu l$,  this condition is violated due to our
choice of fixed parameters that were originally chosen from
published data related to less advanced patients.

\pagebreak

\pagebreak

\begin{figure}
\centerline{\vbox{
\epsfig{file=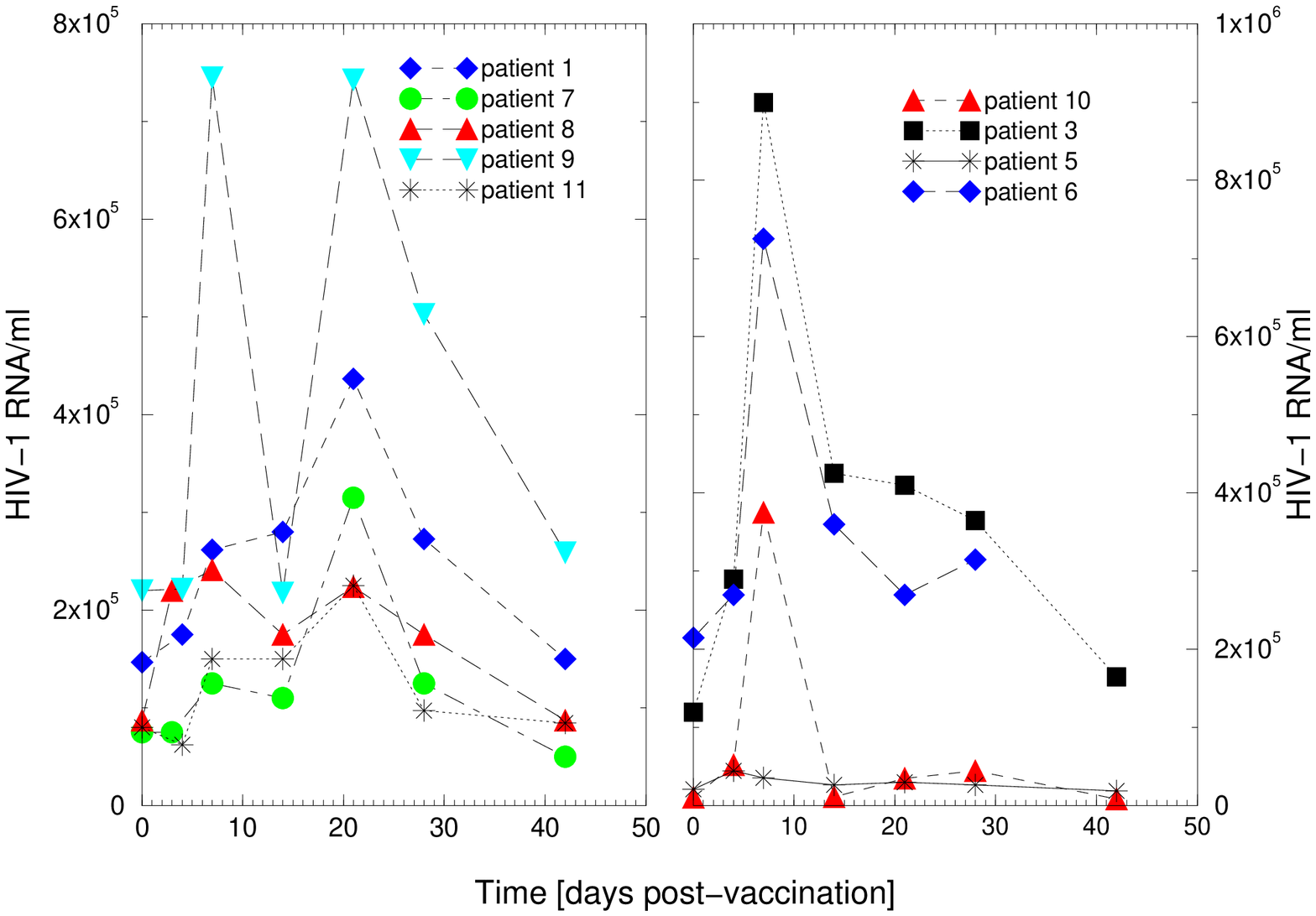,angle=-90,width=5.00in}}}
\caption{ Data from [2] showing post-vaccination rises in
plasma HIV-1 RNA.  (A.) Patients experiencing
two distinct peaks in viremia at roughly days 7 and 21. 
Patients 8 and 9 have equal or nearly equal peaks at days 7 and 21, while patients
1, 7, and 11 experience what was recorded as a 
minor peak at day 7 followed by a true peak at day 21.
(B.) Patients experiencing one early primary peak in viremia at day 7. 
Note that the curves for most of these patients
show a post-peak 'shoulder' at days 21-28.}
\end{figure}

\begin{figure}
\centerline{\vbox{
\epsfig{figure=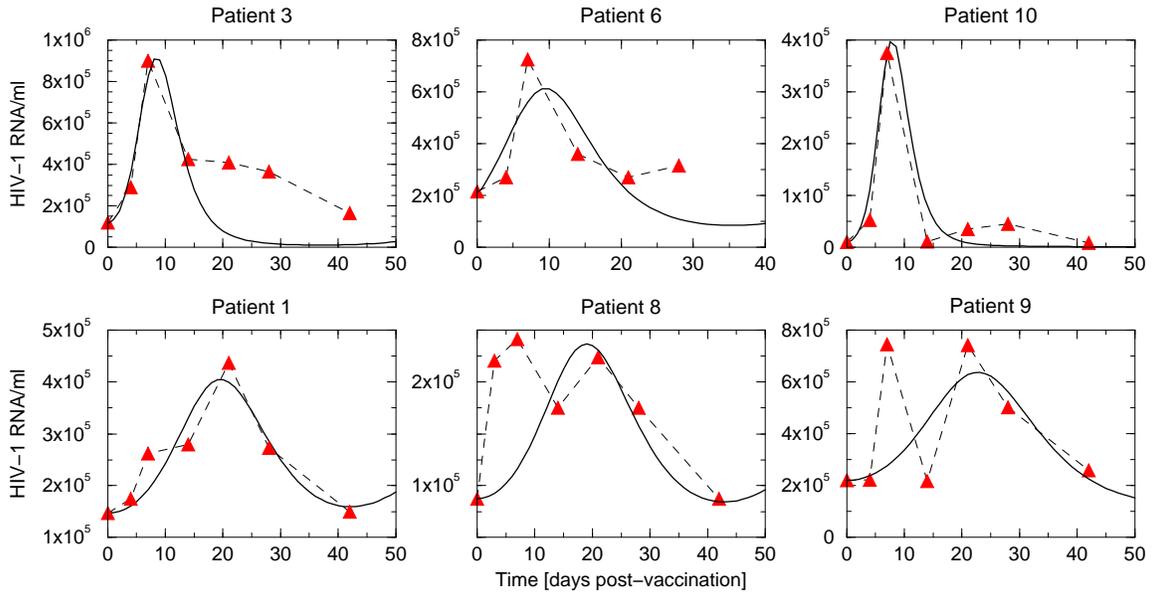,angle=-90,width=6.00in}}}
\caption{ Comparison of  model (solid line) with data
(red triangles).  Note that in some cases data sampling is such that the data curve
is slightly asymmetric, and the true peak may not have been sampled. Modeled
Patient 6 experienced a second very late peak (42 days) associated with known cavitary
pneumonia, so the data is truncated at 28 days.}
\end{figure}

\begin{figure}
\centerline{\vbox{
\epsfig{file=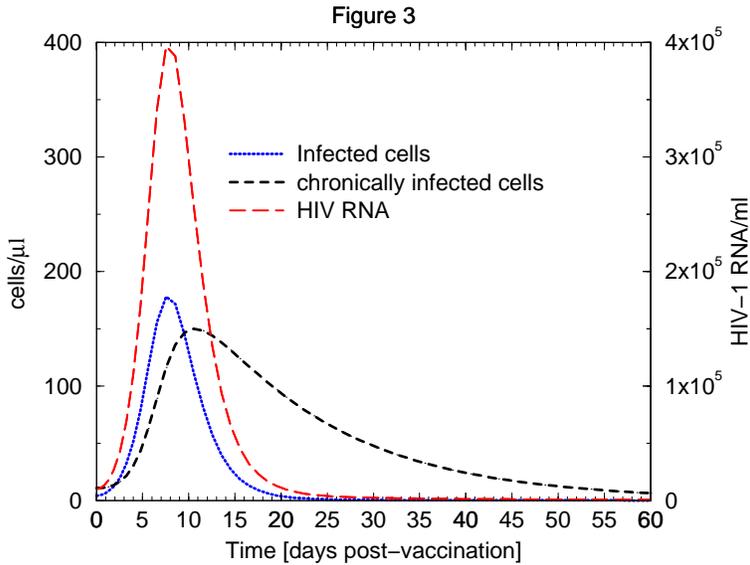,angle=-90,width=4.00in}}}
\caption{Modeled post-vaccination infected cell density and
HIV-1 plasma viral load for patient 10.} 
\end{figure}

\end{document}